\documentstyle[psfig,epsfig]{mn}
\newif\ifAMStwofonts

\ifoldfss
  \ifCUPmtlplainloaded \else
    \NewTextAlphabet{textbfit} {cmbxti10} {}
    \NewTextAlphabet{textbfss} {cmssbx10} {}
    \NewMathAlphabet{mathbfit} {cmbxti10} {} 
    \NewMathAlphabet{mathbfss} {cmssbx10} {} 
  \fi
  \ifAMStwofonts
    \ifCUPmtlplainloaded \else
      \NewSymbolFont{upmath} {eurm10}
      \NewSymbolFont{AMSa} {msam10}
      \NewMathSymbol{\upi}     {0}{upmath}{19}
      \NewMathSymbol{\umu}     {0}{upmath}{16}
      \NewMathSymbol{\upartial}{0}{upmath}{40}
      \NewMathSymbol{\leqslant}{3}{AMSa}{36}
      \NewMathSymbol{\geqslant}{3}{AMSa}{3E}

      \let\leq=\leqslant 
       
    \fi
  \fi
\fi 

\ifnfssone
  \newmathalphabet{\mathit}
  \addtoversion{normal}{\mathit}{cmr}{m}{it}
  \addtoversion{bold}{\mathit}{cmr}{bx}{it}
  \newmathalphabet{\mathbfit} 
  \addtoversion{normal}{\mathbfit}{cmr}{bx}{it}
  \addtoversion{bold}{\mathbfit}{cmr}{bx}{it}
  \newmathalphabet{\mathbfss} 
  \addtoversion{normal}{\mathbfss}{cmss}{bx}{n}
  \addtoversion{bold}{\mathbfss}{cmss}{bx}{n}
  \ifAMStwofonts
    \ifCUPmtlplainloaded \else
      \UseAMStwoboldmath
      \makeatletter
      \new@mathgroup\upmath@group
      \define@mathgroup\mv@normal\upmath@group{eur}{m}{n}
      \define@mathgroup\mv@bold\upmath@group{eur}{b}{n}
      \edef\UPM{\hexnumber\upmath@group}
      \new@mathgroup\amsa@group
      \define@mathgroup\mv@normal\amsa@group{msa}{m}{n}
      \define@mathgroup\mv@bold\amsa@group{msa}{m}{n}
      \edef\AMSa{\hexnumber\amsa@group}
      \makeatother
      \mathchardef\upi="0\UPM19
      \mathchardef\umu="0\UPM16
      \mathchardef\upartial="0\UPM40
      \mathchardef\leqslant="3\AMSa36
      \mathchardef\geqslant="3\AMSa3E

      \let\leq=\leqslant 

    \fi
  \fi
\fi 

\ifnfsstwo
  \DeclareMathAlphabet{\mathbfit}{OT1}{cmr}{bx}{it}
  \SetMathAlphabet\mathbfit{bold}{OT1}{cmr}{bx}{it}
  \DeclareMathAlphabet{\mathbfss}{OT1}{cmss}{bx}{n}
  \SetMathAlphabet\mathbfss{bold}{OT1}{cmss}{bx}{n}
  \ifAMStwofonts
    \ifCUPmtlplainloaded \else
      \DeclareSymbolFont{UPM}{U}{eur}{m}{n}
      \SetSymbolFont{UPM}{bold}{U}{eur}{b}{n}
      \DeclareSymbolFont{AMSa}{U}{msa}{m}{n}
      \DeclareMathSymbol{\upi}{0}{UPM}{"19}
      \DeclareMathSymbol{\umu}{0}{UPM}{"16}
      \DeclareMathSymbol{\upartial}{0}{UPM}{"40}
      \DeclareMathSymbol{\leqslant}{3}{AMSa}{"36}
      \DeclareMathSymbol{\geqslant}{3}{AMSa}{"3E}

      \let\leq=\leqslant 

    \fi
  \fi
\fi 

\ifCUPmtlplainloaded \else
  \ifAMStwofonts \else 
    \def\upi{\pi}
    \def\umu{\mu}
    \def\upartial{\partial}
  \fi
\fi

\title{Dissecting a galaxy: mass distribution of 2237+0305}
\author[C.M. Trott, R.L. Webster]
       {Cathryn M. Trott and Rachel L. Webster\\
	Astrophysics Group, School of Physics, University of Melbourne 3010, Australia}
\date{Accepted 2002 March 27.
      Received 2002 March 13;
      in original form 2001 September 26}

\pagerange{\pageref{firstpage}--\pageref{lastpage}}
\pubyear{0000}

\begin{document}

\maketitle

\label{firstpage}

\begin{abstract}

We determine the mass distribution of a spiral galaxy, 2237+0305
using both gravitational lensing and dynamical constraints. We find that lensing can break the disc-halo degeneracy. 2237+0305 has a sub-maximal disc, contributing 57$\pm$3 per cent of the rotational support at the disc maximum. The disc mass-to-light ratio is 1.1$\pm$0.2 in the I-band and the bulge, 2.9$\pm$0.5. The dark matter halo, modelled as a softened isothermal sphere, has a large core radius (13.4$\pm0.4$ kpc $\equiv$ 1.4$r_d$) to high accuracy for the best-fit solution. The image positions are reasonably well fitted, but require further rotation information to obtain a unique solution.

\end{abstract}


\section{Introduction}

There have been countless papers in recent years discussing the matter distribution in galaxies and, in particular, the contributions of both the stellar disc and the dark matter halo. Any attempt to model the structure of a galaxy with multiple components is limited by the information provided by the light we receive. This introduces degeneracies in the mass distribution when we compare the few constrained parameters with the large number of unknowns.

One technique that has been used to side-step this problem is to assume a maximal disc (quantified in Sackett 1997), whereby the disc contributes the majority of the rotation (75-95\%) at the radius of its maximum circular speed. This definition takes into account the contribution to the inner rotation curve of a bulge component. Various studies have challenged and supported this view and they will be discussed in Section \ref{max}.

Observations of rotation curves of galaxies gave the first hints of the existence of dark matter. In the 1960s and 70s, extensive studies of the circular rotation of spiral galaxies suggested the need for dark matter to account for the missing dynamical mass (e.g. see Rubin et al. 1962; Rubin, Ford \& Thonnard 1978; Faber \& Gallagher 1979). The flatness of rotation curves beyond the optical edge of these spirals contradicted the expected Keplerian fall-off. In the 1980s and 90s, many groups used large scale structure to deduce the expected properties of dark matter particles and applied these to N-body simulations (Peebles 1984; Navarro, Frenk \& White 1996). The subsequent calculated profiles can be compared with the results of observations for consistency.

The firm determination of a general trend in the features of rotation curves, as distinct from a few abnormal systems, requires the accumulation of data from many galaxies with varying morphological types. Salucci \& Burkert (2000) constructed `universal rotation curves' (URCs) with luminosity as the only free parameter, from observations of $\sim$1100 rotation curves. From these URCs, they were able to subtract constant mass-to-light ratio discs from the surface brightness distributions of observed galaxies, and derive the expected contribution from dark matter. Salucci (2001) found the dark matter haloes were required to have large core regions in order to fit the universal curves. Core radii of 3-4 disc scale lengths were found to be consistent with observations for large spiral galaxies.

The use of gravitational lensing to disassemble a galaxy has been used recently by Maller et al. (2000). Their analysis of the doubly imaged system B1600+434 did not have a unique solution due to the necessity of invoking the Tully Fisher relation for rotational information, the large errors on the position angle of the major components and the perturbing effect of a nearby galaxy. In addition, the constraint of only two images limits the number of known parameters. Their work, however, illustrated the utilisation of lensing to break the disc/halo degeneracy. They found a high probability for a sub-maximal disc, and the need for some degree of a constant density core in the centre of the dark matter halo. These results pave the way for our work and demonstrate the ability to solve for the mass distribution given a reasonable number of known parameters.

Galaxy 2237+0305 is improbably close with the source almost exactly aligned with the macrolensing galaxy. Its four, almost symmetric images of the background quasar nicely straddle the central nucleus, placing constraints on the structure of the inner-most regions of the galaxy. Traditionally, it has been difficult to obtain information in these regions due to the degeneracy introduced by multiple mass components and smearing of spectroscopic data from non-circular stellar and gas motions. Lying at low redshift, it displays a visibly extended spiral disc, allowing a detailed investigation of its mass distribution beyond the lensing regions. In addition, two points on the rotation curve have been obtained from neutral hydrogen observations (Barnes et al. 1999). The combination of its proximity and its improbable alignment with the background quasar makes this system the ideal laboratory for studying galactic structure.

Section 2 of this paper discusses the basics of lensing. Section 3 introduces the galaxy 2237+0305 and Section 4 the mass models, our method and the constraints used. Section 5 presents our results and their implications for the structure of the galaxy.

\section{Lensing}

Gravitational lensing can be treated as a purely geometric consequence of General Relativity. If a background source is sufficiently close to the optic axis linking the lens and the observer (approximately within one Einstein radius), the paths of the light rays received from it will have been altered by the mass of the lensing object. If the simplest case of point masses for both lens and source is assumed, the positions of the images observed in the lens plane can be obtained analytically using simple geometry. This geometry is directly generalisable for more complex mass distributions, where extended lenses comprise multiple components. The lens equation linking these can be written easily,

\begin{equation}
\bmath{\alpha}(\bmath{\theta}) = \bmath{\theta} - \bmath{\beta},
\end{equation}
where

\begin{equation}
\bmath{\alpha} = \frac{D_{ds}}{D_s}\bmath{\hat{\alpha}},
\end{equation}
is the reduced bending angle, $\bmath{\theta}$ is the image position, $\bmath{\beta}$ is the source position, and all distances are angular diameter distances (Narayan \& Bartelmann 1999). Throughout, we have assumed an Einstein-de Sitter universe with a Hubble constant of $H_0$ = 70 kms$^{-1}$Mpc$^{-1}$, although cosmology makes little difference to physical scales of the galaxy due to its proximity to the observer.

The angle of deflection of a light ray travelling from source to observer is dependent upon the projected surface mass distribution in the lens plane of the lensing object. This incorporates both the convergence (the greater the surface mass density, the greater the deflection) and shear (the greater the anisotropy of the lens, the greater the deviation of the images from a uniform, circular configuration) of the lens. The bending angle, $\bmath{\hat{\alpha}}$ is given by,

\begin{equation}
\bmath{\hat{\alpha}}(\bmath{\xi})=\frac{4G}{c^{\rm 2}}{\int}\frac{(\bmath{\xi}-\bmath{\xi}^\prime)\Sigma(\bmath{\xi}^\prime)}{|\bmath{\xi}-\bmath{\xi}^\prime|^{\rm 2}}d^{\rm 2}\xi^\prime,
\end{equation}
where $\Sigma$ is the surface mass density and the integral extends over the surface of the distribution (primed co-ordinates). The addition of multiple mass components is trivial with this formalism and allows a complicated system to be applied to lensing.

\section{2237+0305}

2237+0305, a barred spiral Sab type galaxy at a redshift of $z$=0.0394 was first discovered by Huchra et al. (1985), but the images were first seen by Yee (1988). Almost collinear with the centre of the galaxy (the optic axis) is a background quasar at $z$=1.695. The geometric conditions of this system are favourable to macrolensing of the quasar through the bulge of the galaxy.

\begin{figure}
\begin{center}
 \epsfig{file=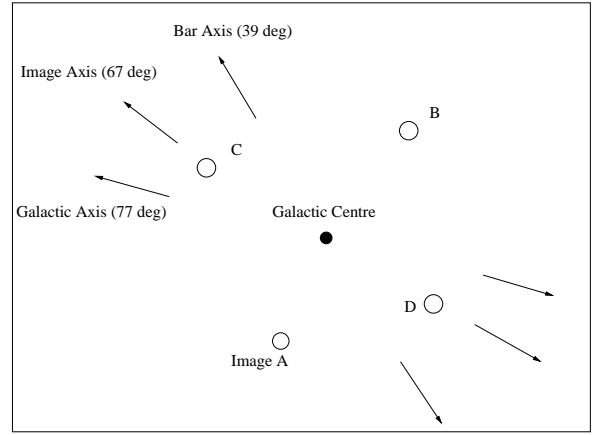,width=75mm,angle=270}
 \caption{Locations of the four lensed images as measured by Crane et al. (1991) and the major axes of the principal galactic components as measured by Yee (1988).}\label{images}
\end{center}
\end{figure}

Huchra's lens, as it is often termed, has been extensively studied over the past fifteen years, mainly due to its proximity to us, and the ability to image the galaxy with ground-based telescopes. For this reason, it is the ideal laboratory in which to study lensing, and more importantly for this work, the structure of the lensing galaxy. The lens has a visible central bulge, a stellar disc and a perturbing bar. The four images straddle the nucleus in an Einstein Cross configuration and their positions have been measured accurately (Crane et al. 1991). The major axis of the disc and bulge have been measured by Yee at a position angle of 77$^{\circ}$, and the images at an angle of 67$^{\circ}$. The rotation of the images away from the major axis of the galaxy is due to the torqueing effect of the bar at a position angle of 39$^{\circ}$. Schmidt et al. (1998) extensively modelled the mass distribution in the bar and its effect on the positions of the images. The angular positions of the major components of the galaxy are displayed in Figure \ref{images}.

The scale lengths of the luminous parts of the galaxy have been measured from HST light profiles by Schmidt (1996), and their ellipticities by Racine (1991) and Irwin et al. (1989), and thus many of the galactic parameters have been previously measured.

Qualitatively, the majority of the convergence will be provided by the central bulge (in the absence of a cuspy dark matter halo) as it dominates in the region in which the lensing is most affected. The disc will contribute both globally and add significant convergence to the inner regions. The bar will add little convergence for the lensing, but will provide shear through its rotated position angle. In addition, the inclination of the disc and flattening of the bulge will also contribute to the twisting of the image positions. The presence of a dark matter component to the galaxy will add convergence and, if non-spherical, shear. Given the highly elliptical nature of the visible components already discussed, we will assume a spherical halo.

\section[]{Models}

\subsection{Mass Models}

We will construct models for the mass distributions of the bulge, disc, bar and dark matter halo using parameters from previous studies, and map light rays from the measured image positions to the source plane. Varying the contribution of each component will vary the convergence and shear within the images and shift their back-mapped positions in the source plane. A potential solution is obtained when a particular addition of the four components produces a common source position. Clearly, the four images originate from the same point in the source plane. The actual position of the source quasar cannot be constrained. The centre of the galaxy will be considered fixed given the relatively few degrees of freedom.

Rotation curves for potential solutions will be produced and compared with neutral hydrogen rotation measurements and the measured mass lying within the images. The combination of both lensing and dynamical constraints increases the number of constrained parameters and consequently reduces the number of unknowns.

The models used for the mass distributions of the four principal galactic components are standard profiles from the literature, tailored to suit this galaxy.

The bulge is modelled as both a modified de Vaucouleurs surface mass distribution (de Vaucouleurs 1948, 1959), where it is assumed the mass follows the light (constant mass-to-light ratio) and an exponential surface mass profile, as in the models of Schmidt (1996). The modification to both allows the introduction of an ellipticity, $e$, such that for the de Vaucouleurs profile,

\begin{equation}
\log\Sigma({x},{y}) = \log{\Sigma_0}-3.33\left[\left(\frac{x^2 + \frac{y^2}{(1-e)^2}}{r_b}\right)^\frac{1}{4}-1\right],
\end{equation}
where $\Sigma$ is the value of the surface mass density at that position, $r_b$ is the characteristic scale length of the bulge, and $e$ is defined by,

\begin{equation}
e = 1 - \frac{b}{a},
\end{equation}
where $a$ and $b$ are the semi-major and minor axes respectively. The central surface density ($\Sigma_0\times$10$^{3.33}$) is denoted `bg'. The exponential profile is modelled simply by introducing the ellipticity (assumed to be a projection effect),

\begin{eqnarray}
\Sigma({x},{y}) = {\rm bg}e^{-f(x,y)/r_b},\\
f(x,y) = \sqrt{x^2+\frac{y^2}{(1-e)^2}}
\end{eqnarray}
and again `bg' is the central surface mass density.

The disc is modelled as an exponential surface density. Unlike the bulge which is treated with the ellipticity as measured, the disc is rotated to its measured inclination of $i$=60$^{\circ}$. This involves projecting the volume to a surface mass density by rotating the $z$-axis and redefining co-ordinates. If we assume the disc is uniformly distributed in the $z$ direction, then we can simply write,

\begin{equation}
\rho(x,y,z) \propto e^{-\sqrt{x^2+y^2}/r_d},
\end{equation}
where the proportionality includes a factor reflecting the thickness of the disc, assumed to be constant, and $r_d$ is the characteristic disc scale length. Upon rotation about the $x$-axis (such that it becomes the major axis of the ellipse) by the inclination angle, $i$=60$^{\circ}$, the surface mass density is the integral through the rotated axis, $z^\prime$,
\begin{equation}
\Sigma(x^\prime,y^\prime) = {\rm dc} {\int\limits_{z_{min}^\prime}^{z_{max}^\prime}}e^{-\sqrt{x^{\prime2} + ((z^\prime-z_{min}^\prime)\sin{i} + \frac{y^\prime}{\cos{i}})^2}/r_d}\,dz^\prime,
\end{equation}
where the limits of integration bound the original constant disc thickness at the inclination angle (taken as $\Delta{z}$=500pc), the primed co-ordinates represent the new, observed Cartesian system and `dc' denotes the central surface mass density.

The bar has been extensively modelled by Schmidt (1996) and we will use his surface mass distribution and position angle. Schmidt uses a Ferrers model with an ellipticity, $e$,

\begin{equation}
\Sigma(x,y) = {\rm br}\left(1 - \frac{x^2}{a^2} - \frac{y^2}{b^2}\right)^\lambda,
\end{equation}
where `br' is the central surface density, $\lambda$ is the Ferrers exponent, and the $(x,y)$ co-ordinates lie in the rotated frame of the bar. Schmidt finds different exponents, ellipticities and scale lengths depending on the profiles used to fit to the light distribution. For an exponential bulge and disc, he finds $\lambda$=2, $e$=0.64 and b=1.0$\pm$0.3 arcsec fit the observations best. For a de Vaucouleurs bulge and exponential disc, he finds $\lambda$=0.5, $e$=0.89 and b=3.1$\pm$0.9 arcsec. In our analysis, the central surface density (essentially the M/L) will remain a free parameter. Here $\Sigma_{cr}$ is the critical surface density defined by the geometry,

\begin{equation}
\Sigma_{cr} = \frac{c^2}{4{\pi}G}\frac{D_s}{{D_d}D_{ds}},
\end{equation}

The profile of the dark matter halo is controversial. Conventional theories where $\Lambda$CDM is the preferred cosmology have been very successful in explaining the observed large scale structure of the universe (eg Peebles 1984). Navarro, Frenk and White (1996; hereafter NFW) used N-body simulations to derive a density profile for such a cosmology. One feature of the NFW profile is a cuspy central region with $\alpha \sim$ -1 where $\rho \sim r^{\alpha}$. This steepens to $\alpha \sim$ -3 for $r{\gg}r_h$, where $r_h$ is the characteristic scale length of the halo. Recently, more accurate simulations have pushed the cuspiness at $r \sim$ 0 to $\alpha \sim$ -1.5 (eg Moore et al. 1999a). Recent observational work by de Blok et al. (2001) on low surface brightness galaxies has shown the need for a core in the dark matter haloes, using optical rotation curves and fitting minimal discs. CDM has been challenged further by the missing satellite (Moore et al. 1999b) and angular momentum problems (see recent discussion by Sommer-Larsen \& Dolgov 2001). The former considers the few satellite galaxies observed in orbit around our Galaxy compared with that predicted by the theory, and the latter refers to the predictions of CDM of too much angular momentum loss to support observed disc galaxies.

Since there is no general agreement we have chosen to model the dark matter halo as a softened isothermal sphere (Kormann, Schneider \& Bartelmann, 1994),

\begin{equation}
\Sigma(r) = \frac{\sigma_v^2}{2G}\frac{1}{\sqrt{r^2+r_c^2}},
\end{equation}
where $\sigma_v$ is the velocity dispersion and $r_c$ is the core or break radius. There are several reasons for this choice of profile. Firstly, it has one less parameter than an NFW+core profile (no scale length as distinct from the core length). Given the small number of degrees of freedom we have, an extra one is useful. Secondly, the profile naturally asymptotes to a flat rotation curve for $r \gg r_c$, and finally, recent results find that the two models are indistinguishable when the remaining mass components are taken into account (Weiner, Sellwood \& Williams, 2001, modelled NGC 4123 with both profiles and found their shape had a minor effect on the results). In addition to this, the choice of a spherical halo is a simplification in that it precludes the need to introduce another parameter. This choice is partially justified by the ellipticities in the other components. They are adequate to produce the required ellipticity. Furthermore, recent results from Ibata et al. (2001) suggest dark matter haloes are spherical. They used evidence of the tidal stream from the Sagittarius dwarf galaxy to show that the halo potential cannot be flatter than $q<0.7$ and probably has $q>0.9$, where $q$ is the axis ratio.

\subsection{Constraints and Previous Work}

High resolution imaging and detailed modelling of Huchra's Lens has provided us with many of the model parameters required for fitting a mass distribution. Schmidt (1996) used I-band \textit{HST} imaging to measure the disc and bulge scale lengths, as well as to determine their ellipticities. These results were updated values from previous studies by Yee (1988) and Huchra et al. (1985). These data leave the disc and bulge mass-to-light ratios as the only fitting parameters for these components. The bar has a measured ellipticity, position angle and major axis, and within mass-to-light ratio uncertainties is completely determined by Schmidt.

Schmidt deconvolved the \textit{HST} light distribution using two different bulge models, an exponential and a de Vaucouleurs surface mass profile. Results from the literature show that one of these profiles commonly fits a bulge well. Carollo et al. (2001) show from \textit{HST} observations that the bulge mass profile is mildly morphologically dependent. For Sa-Sb galaxies, such as 2237+0305, both types are observed and thus an exploration of both models is prudent.

One further constraint and one check will be applied to the final rotation curve. Neutral hydrogen observations by Barnes et al. (1999) at the VLA provide two rotation points in the outer regions of the galaxy. In these regions, the visible matter has fallen below observational levels and the HI is acting as a tracer of the dark matter distribution. Unfortunately, the data are not of high enough angular resolution to probe the rotation in the inner regions of the galaxy. An additional piece of information is provided by gravitational lensing, where the position of the images in combination with the geometry of the source-lens-observer system gives the projected mass enclosed within the images (Rix, Schneider \& Bahcall, 1992; Wambsganss \& Paczynski, 1994). The consistency of this value with the mass distribution found will act as a check on the result. We define the image radius as the average of the radii of the four images from the centre of the galaxy. This corresponds to $r_{im}$ = 0.9 arcsec $\equiv$ 670h$_{70}^{-1}$ parsecs. The halo is completely unconstrained observationally. We will fit to both the scale length and the halo normalisation.

We therefore have seven parameters for which a fit is required - four mass-to-light ratios, one core radius and two co-ordinates of the source position. We have as constraints, eight co-ordinates of image positions, two rotation points and one mass enclosed within the images. These observational constraints are given in Table \ref{values}. Thus, we have four degrees of freedom.
\begin{table}
\centering
\begin{tabular}{l|l|l}	\hline
{\em Quantity} & {\em Value} & {\em Source}\\	\hline\hline
{\em Halo} & \\	\hline
No constraints\\ \hline
{\em Bulge} & \\	\hline
Position angle & 77$^\circ$ & Yee (1988)\\
Ellipticity & 0.31 & Racine (1991)\\
{\bf dV} & \\
Scale length & 4.1$\pm$0.4$''$ (3.1kpc) & Schmidt (1996)\\
{\bf Exp} & \\
Scale length & 0.59$\pm$0.03$''$ (0.45kpc) & Schmidt (1996)\\ \hline
{\em Disc} & \\	\hline
Position angle & 77$^\circ$ & Yee (1988)\\
Inclin. angle & 60$^\circ$ & Irwin et al. (1989)\\
{\bf dV} & \\
Scale length & 11.3$\pm$1.2$''$ (8.6kpc) & Schmidt (1996)\\
{\bf Exp} & \\
Scale length & 5.6$\pm$0.4$''$ (4.27kpc) & Schmidt (1996)\\ \hline
{\em Bar} & \\	\hline
Position angle & 39$^\circ$ & Yee (1988)\\
{\bf dV} & \\
Sfc. Brightness & $I_0$=20.4$\pm$0.2 & Schmidt (1996)\\
Ellipticity & 0.64 & \\ 
{\bf Exp} & \\
Sfc. Brightness & $I_0$=19.9$\pm$0.2 & Schmidt (1996)\\
Ellipticity & 0.89 & \\ \hline
{\em Images} & \\	\hline
Mass Enclosed & (1.48$\pm$0.01) h$_{75}^{-1}$ & WP (1994)\\
($\times$10$^{10}$M$_\odot$) & = (1.59$\pm$0.01) h$_{70}^{-1}$ & \\	\hline
Positions ('') & $\Delta$RA $\:$ $\Delta$Dec. & Crane et al. (1991)\\
Image A &  0.093 $\:$ -0.936 & \\
Image B &  0.579 $\:$ 0.737 & \\
Image C & -0.719 $\:$ 0.266 & \\
Image D &  0.761 $\:$ -0.419 & \\	\hline
{\em Rotation:$v_{circ}$} & \\	\hline
$22\pm1$ kpc & 310$\pm$15 kms$^{-1}$ & Barnes et al. (1999)\\
$29\pm1$ kpc & 295$\pm$15 kms$^{-1}$\\
\end{tabular}\caption{Assumed values for the mass distributions from previous work. Image positions are relative to the galactic centre, and have positional uncertainties of $\pm$5mas. WP (1994) denotes Wambsganss \& Paczynski (1994), and the central I-band magnitudes ($I_0$) are measured in mag./arcsec$^2$. {\bf dV} and {\bf Exp} denote models with a de Vaucouleurs and an exponential bulge, respectively.}\label{values}
\end{table}

The galaxy has been previously modelled by many groups. Huchra et al. (1985) undertook the initial work on the system, measuring ellipticities and scale lengths and providing a rudimentary lensing analysis. They assumed a circular lens model and showed that the inferred mass-to-light ratio is within current values for nearby galaxies. Kent \& Falco (1988) approximated the galaxy as an oblate spheroid, citing the bulge and the bar as the two primary lensing components. Their analysis attempted to fit the observed quasar fluxes to their model and was reasonably successful. Contemporaneously, Schneider et al. (1988) used a single, elliptical de Vaucouleurs bulge to model the galaxy mass distribution. They fitted two free parameters, the mass-to-light ratio (assumed to be constant) and the source position, given the image positions, by minimising the deviation of predicted positions to those observed. Unfortunately, they were limited by imaging taken with large PSFs and by the simplicity of their modelling. Both Kent \& Falco, and Schneider's group find different positions for the background source, as does Schmidt (1996) in his bar-centric analysis. Schmidt combines the disc and bulge into a single component with the ellipticity Racine (1991) measured for the bulge alone. The bar was then studied in detail using high resolution \textit{HST (WF/PC-1)} I-band imaging and the source position again back-mapped from the measured image positions. This is the first study to include the dark matter halo in the modelling.

In order to have a reasonable chance of understanding the complex structure of this lens, and hence its lensing characteristics, one needs to model each component of the lens carefully. With the accumulation of data on this galaxy over the past fifteen years, this is possible.

\section{Results}

\subsection{Source Position and Rotation Curve}

The lensing galaxy was constructed as a two-dimensional array of surface mass elements, with an exponential spatial scale. This feature allowed the important central regions to be highly resolved, while keeping the array dimensions computationally tractable. The galaxy was rotated from its position angle to make the $x$-axis coincident with the galactic major axis, and the image positions rotated to a right-handed co-ordinate system. The four mass components were constructed and overlayed on the mass array, forming a surface mass distribution for the galaxy. Alterations were made to this array when varying the halo core radius, $r_c$ or the scale lengths of the bulge and disc. The bending angles for the particular distribution were then calculated.

We have calculated bending angles for particular values of bulge and disc scale length, for a given bulge profile. These values for the four components then were added in linear combination varying the mass-to-light ratios of each component. Since the bending angles scale linearly with mass-to-light ratio, variations in the overall scaling parameters can be considered after the angles have been calculated.

Rotation curves were carefully constructed to incorporate the lack of spherical symmetry in the disc and bulge. The functional form for the disc rotation was obtained for an exponential disc from Binney \& Tremaine (1987). The de Vaucouleurs bulge was de-projected into an elliptical volume mass density using the profile of Fillmore (1986) and the rotation calculated using mass enclosed within a radius and equating the centripetal and gravitational accelerations.

The mass measured by Wambsganss \& Paczynski (1994) within the images uses the relations of lensing and determines the projected mass in a cylinder bounded by the images and integrated along the line-of-sight. A suitable distribution of mass profiles must satisfy this constraint. The consistency between this mass and a solution that reproduces the correct image positions does not provide a completely independent constraint as it has been derived from lensing. It only demonstrates the true convergence and shear that has already been accounted for in the source finding program. As such, the consistency of these points provides a check that the source finding program is reconstructing the correct shear and convergence.

The process of deciding which distributions adequately fit all of the constraints reduces to the minimisation of the $\chi^2$ statistic. For a particular set of profiles for the luminous mass components (as given in Table \ref{values}), a search through five dimensional parameter space (four mass-to-light ratios and one core radius) was performed. For each model providing a suitable mass enclosed within the images, the eight image positions and the two points at the pertinent positions on the rotation curve, were calculated. These were used to calculate the $\chi^2$ statistic and the minimum found. In addition to the variation of the mass profiles (de Vaucouleurs ({dV}) versus exponential ({Exp}) bulge, and subsequent changes in disc and bar profiles), the scale lengths of the disc and bulge were varied to their 1$\sigma$ observational limits in order to obtain a more complete representation of the parameter space.

The incomplete Gamma Function provides the goodness of fit for a system with $n$ degrees of freedom (d.o.f.) and a $\chi^2$ statistic. For a system with four d.o.f., a solution with the probability to 1$\sigma$ (68\%) of being real (that is, not by chance) is given by solving,

\begin{equation}
1 - F(n/2;\chi^2/2) = 0.68
\end{equation}
where $F(\alpha;\beta)$ is the incomplete gamma function (Numerical Recipes). The solution for $n$=4 is $\chi^2_{\rm max}$=2.3. This is a strict value from the formal theory and requires a high degree of accuracy to be attained. Considering the idealistic mass models we are using and the small number of d.o.f., we would not expect to reach this low value for the $\chi^2$. Instead, the minimum value found will be taken as an adequate fit, provided it does not exceed this value excessively. The error analysis for the best fit result will provide information on the sensitivity of the solution on the different parameters. We will take the errors on a parameter to be where the $\chi^2$ increases by one while holding all other parameters constant.

The results for the model with a de Vaucouleurs bulge is shown in Table \ref{param_dev}, and an exponential bulge in Table \ref{param_exp}.
\begin{table}
\centering
\begin{tabular}{l|c|c|c|c|c}	\hline
{\em $r_b$} & {3.1} & {3.1} & {3.1} & {2.8} & {3.4} \\
{\em $r_d$} & {8.6} & {9.5} & {7.7} & { 8.6} & {8.6} \\  \hline
{\em $r_c$} & {15.3$\pm$0.4} & {13.4$\pm$0.4} & {14.9$\pm$0.5} & {-} & {14.5$\pm$0.4} \\
{\em $\sigma_v$} & {266$\pm$4} & {233$\pm$4} & {246$\pm$4} & {-} & {257$\pm$4} \\
{\em bg} & {9.66$\pm$0.04} & {9.53$\pm$0.04} & {9.47$\pm$0.04} & {-} & {8.66$\pm$0.04} \\
{\em dc} & {242$\pm$24} & {506$\pm$30} & {571$\pm$38} & {-} & {447$\pm$35} \\
{\em br} & {824$\pm$23} & {821$\pm$20} & {818$\pm$20} & {-} & {783$\pm$22} \\  \hline
{\em $\chi^2$} & {10.95} & {5.26} & {6.99} & {$>$500} & {9.81} \\
\end{tabular}\caption{Best-fit parameters for a given model, exponential disc and de Vaucouleurs bulge. The scale lengths ($r_b$, $r_d$) are given in kpc; {\em $\sigma_v$} is the velocity dispersion (km/s) of the halo profile; {\em bg}, {\em dc} and {\em br} are the central densities (M$_\odot$pc$^{-2}$) for the bulge, disc and bar respectively. The bulge values shown are divided by 10$^5$ of their actual value.}\label{param_dev}
\end{table}

\begin{table}
\centering
\begin{tabular}{l|c|c|c|c|c}	\hline
{\em $r_b$} & {0.45} & {0.45} & {0.45} & {0.43} & {0.47} \\
{\em $r_d$} & {4.27} & {4.58} & {3.97} & {4.27} & {4.27} \\  \hline
{\em $r_c$} & {15.2} & {15.2} & {15.2} & {15.9} & {15.7} \\
{\em $\sigma_v$} & {243} & {243} & {243} & {262} & {239} \\
{\em bg} & {3.0$\times$10$^4$} & {3.0$\times$10$^4$} & {3.0$\times$10$^4$} & {3.2$\times$10$^4$} & {2.9$\times$10$^4$} \\
{\em dc} & {0.0} & {0.0} & {0.0} & {3.6} & {0.0} \\
{\em br} & {924} & {924} & {924} & {934} & {894} \\  \hline
{\em $\chi^2$} & {140} & {140} & {140} & {106} & {210} \\
\end{tabular}\caption{Same as for Table \ref{param_dev} except for an exponential disc and bulge. Symbols have the same meaning, but the bulge scaling is correct. Uncertainties were not calculated due to the large $\chi^2$ values.}\label{param_exp}
\end{table}
These tables clearly demonstrate the superiority of the former over the latter. The $\chi^2$ parameter for the exponential bulge results is evenly distributed between error in the rotation curve points and the image positions. It seems the rotation curve cannot be sufficiently produced to satisfy the mass enclosed within the images constraint as well as the two points measured in the curve. Similarly, the combination of bulge, disc and bar cannot produce the required ellipticity as well as an acceptable rotation curve.

Conversely, the results for the de Vaucouleurs bulge and exponential disc are quite promising. None of the combination of scale lengths produced an acceptable $\chi^2$ ($\leq$2.3), however, the values are of the right order of magnitude. The best result is produced at the 1$\sigma$ upper limit of the disc scale length as measured by Schmidt. Without a continuity of solutions between these scale lengths, it is difficult to predict if an acceptable solution occurs within the measured limits. Without the added accuracy of further rotation points, we will take this solution to be acceptable and base our analysis upon it. The $\chi^2$ for this result is distributed evenly between error in the rotation points and the image positions.

\begin{figure*}
\begin{center}
\epsfig{figure=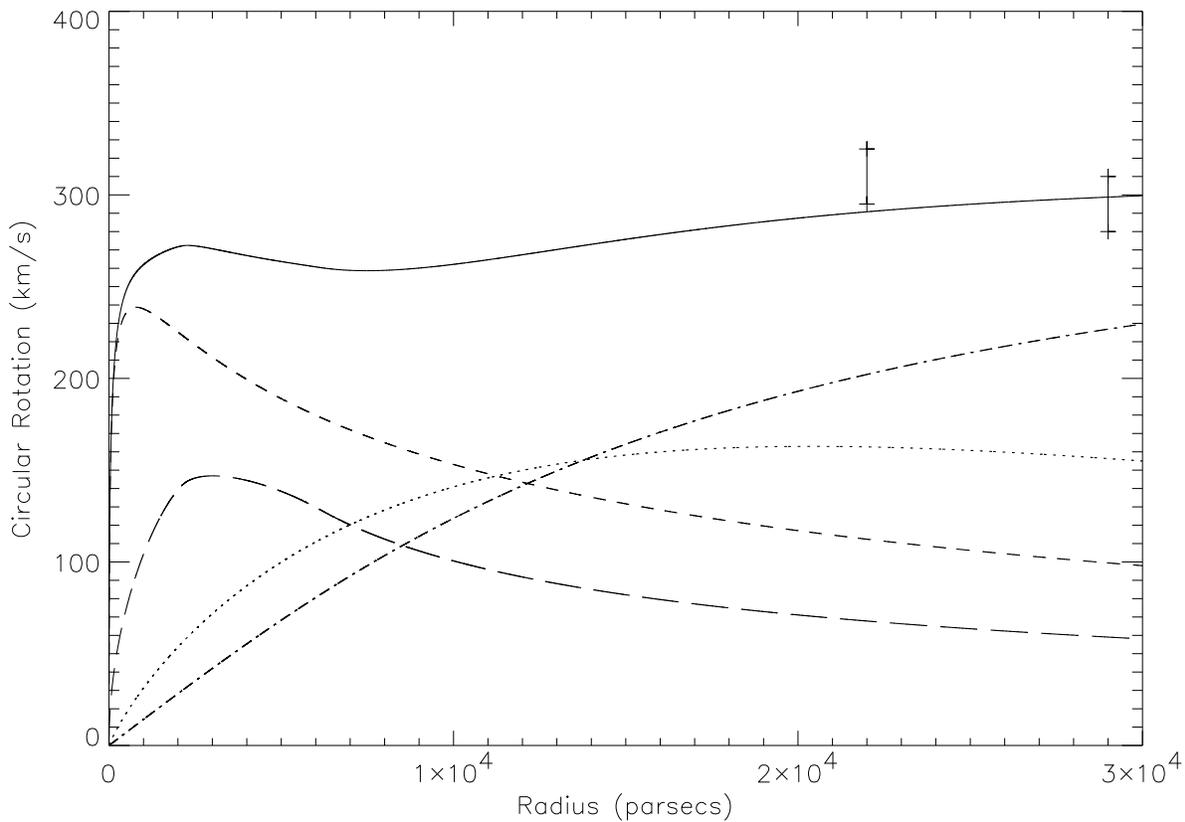,width=165mm}
\caption{Rotation curve for the best solution ($r_b$=3.1kpc, $r_d$=9.5kpc, $r_c$=13.4kpc, $\sigma_v$=233km/s, bg=9.53$\times$10$^5$M$_\odot$/pc$^2$, dc=506M$_\odot$/pc$^2$, br=821M$_\odot$/pc$^2$) as in Table \ref{param_dev}. The curves are total (solid), halo (dash-dot), disc (dotted), bulge (short dashed) and bar (long dashed). The two vertical lines denote the rotation points from HI measurements and their 1$\sigma$ uncertainties.}\label{rot_curve}
\end{center}
\end{figure*}

Figure \ref{rot_curve} shows the rotation curve produced from this solution and the components used to construct it.
The total rotation curve is still rising at the edge of the plot due to the influence of the halo. Here, the three luminous components are falling away, particularly the bulge and bar which are essentially contributing no mass. This rise could pose a problem in that rotation curves are observed to be reasonably flat or slowly falling. Simply scaling down the halo core size would alleviate this problem, however it would also affect the convergence at the image positions. Although the halo does not contribute significant mass within the images, losing this would require a scaling up of another component. As all of the other modelled components contribute shear, this would effect the overall image positions and increase the $\chi^2$. Another halo profile may provide a better fit.

 As mentioned above it is pertinent to calculate the projected mass of each of the components within the images. This involves the integration of mass within a cylinder out to the image radius and through the line-of-sight. Table \ref{mass_enc} displays the results.
\begin{table}
\centering
\begin{tabular}{c|c}	\hline
{\em Component} & {\em Mass Enclosed ($\times$10$^{\em{8}}$M$_\odot$)}\\	\hline\hline
{\em Halo} & 6.8$\pm$0.2 \\
{\em Bulge} & 130.3$\pm$0.5 \\
{\em Disc} &  7.1$\pm$0.4 \\
{\em Bar} &  11.8$\pm$0.3 \\	\hline
{\em Total} & 156.0$\pm$1.4 \\
\end{tabular}\caption{Projected mass within the image radius divided into the four mass components for the best fit solution of Table \ref{param_dev}.}\label{mass_enc}
\end{table}
The bulge clearly dominates the convergence within the image region. A percentage variation in the overall scalings of the halo, disc and bar components will make less difference to the lensing than a change in the bulge value. These results demonstrate the nature of stacking these mass components. An increase in the halo mass (for example, due to a small core region) would necessitate a decrease in the bulge. For this to be an adequate solution, however, the halo cannot be too large, otherwise it contributes too much rotation at the HI points, and the loss in shear from the bulge must be compensated for in another elliptical component. Thus, the combination of requiring correct convergence \textit{and} shear is able to break the disc/halo degeneracy, removing the need to consider both maximal and minimal discs (Maller et al. 2000).

In addition to finding the best-fitting bulge and disc contributions to the rotation curve, one can perform a similar analysis on the source position for each potential solution. The results of this analysis are shown in Table \ref{source_posn}, compared with those from previous groups. The source position for a given configuration was found by averaging the backmapped positions of the four images.
\begin{table}
\centering
\begin{tabular}{l|c|c}	\hline
{\em Group} & {\em $\Delta$N (arcsec)} & {\em $\Delta$E (arcsec)}\\	\hline\hline
1. Kent \& Falco (1988) & -0.02$\pm$0.01 & -0.08$\pm$0.02\\
2. Schneider et al. (1988) & 0.015$\pm$0.005 & -0.004$\pm$0.005\\
3. Schmidt et al. (1998) & -0.014$^{-0.003}_{+0.001}$ & -0.063$^{-0.010}_{+0.009}$\\
4. This work & -0.014$\pm$0.001 & -0.072$\pm$0.001\\
\end{tabular}\caption{Mean and standard deviation of the source position for different groups, as measured from the centre of the galaxy. $\Delta$N and $\Delta$E refer to the offset from the centre of the galaxy in the north and east directions.}\label{source_posn}
\end{table}
Our results fit well with those of Schmidt et al. (1998) and Kent \& Falco (1988), but not with Schneider et al. (1988). The consistency of three of the results is encouraging.

\begin{figure}
 \begin{center}
\epsfig{figure=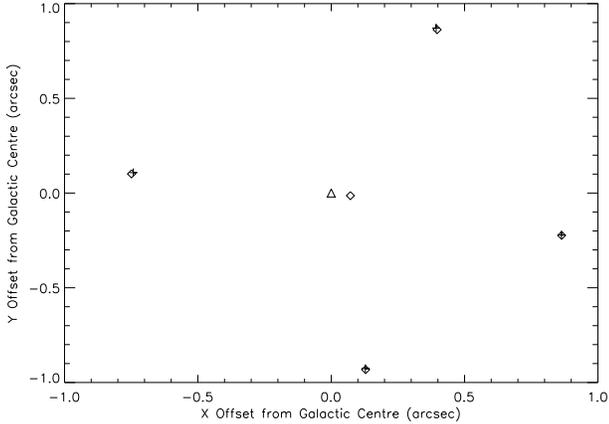,width=85mm}
 \caption{Comparison of the model and measured image positions for the best-fit solution. The crosses denote the model positions, the adjacent diamonds the measured, the triangle is the assumed galactic centre, and the isolated diamond is the source position. The 1$\sigma$ uncertainties on the measured image positions correspond approximately to the size of the diamonds. The positions are labelled A, C, B, D clockwise from the bottom using the Yee (1988) convention.}\label{source_posn_diag}
\end{center}
\end{figure}

The image positions reproduced with this model are not adequate to definitively conclude this as a solution, however they are still extremely close to the observed positions. Figure \ref{source_posn_diag} displays the model and real image positions, as well as the location of the source.

The final piece of relevant information that can be derived from these results is the total mass of each of the galactic components. The optical disc extends out to $r \sim$ 25-30 kpc (Yee 1988) and the dark matter halo will presumably continue far beyond this radius. As such, the masses calculated will be those enclosed within a sphere with radius $r$ = 30 kpc, centred on the galactic nucleus. The results are displayed in Table \ref{total_mass}.
\begin{table}
\centering
\begin{tabular}{c|c}	\hline
{\em Component} & {\em Mass ($<$30 kpc) h$_{70}^{-1}$M$_\odot$}\\	\hline\hline
Halo & (3.7$\pm$0.1) $\times$ 10$^{11}$\\
Bulge & (6.7$\pm$0.1) $\times$ 10$^{10}$\\
Disc & (1.7$\pm$0.1) $\times$ 10$^{11}$\\
Bar & (2.3$\pm$0.1) $\times$ 10$^{10}$\\  \hline
Total & (6.2$\pm$0.2) $\times$ 10$^{11}$\\
\end{tabular}\caption{Masses of the four galactic components from the potential solutions of Table \ref{param_dev}. The values given are total mass within a radius of 30 kpc in solar masses. The range of halo values demonstrate the variation in total halo mass for the potential solutions.}\label{total_mass}
\end{table}
 These values are consistent in their order of magnitude with nominal masses for galactic components. The dark matter halo contributes $\sim$60 per cent of the dynamical mass of the galaxy within the optically visible region. If the halo was allowed to extend further, as it clearly does from its rotation curve, it would contribute much more to the overall galactic mass.

\subsection{Mass-to-Light Ratio}

In order to facilitate his analysis of the bar in 2237+0305, Robert Schmidt decomposed light profiles by taking cuts along the major and minor galactic axes. These profiles can be used to investigate the mass-to-light ratio of the bulge and disc in the I-band (Schmidt, private communication). A deviation from a constant mass-to-light ratio across an individual component could either indicate the non-applicability of the mass profiles to this galaxy, or an intrinsic colour variation. Such a failure of the light to trace the mass would have implications for analyses of galactic discs where the assumption of constant mass-to-light ratio is often made.

The major axis light profile, once calibrated to units of solar luminosity, clearly displays the addition of two distinct components. The central region is dominated by a steeply falling bulge region, which flattens to a more gradually decreasing disc. The luminosity was calculated assuming a solar absolute magnitude in the V-band and correcting it to the I-band using the Vilnius spectral shifts of Bessell (1990).

\begin{figure}
\begin{center}
 \epsfig{file=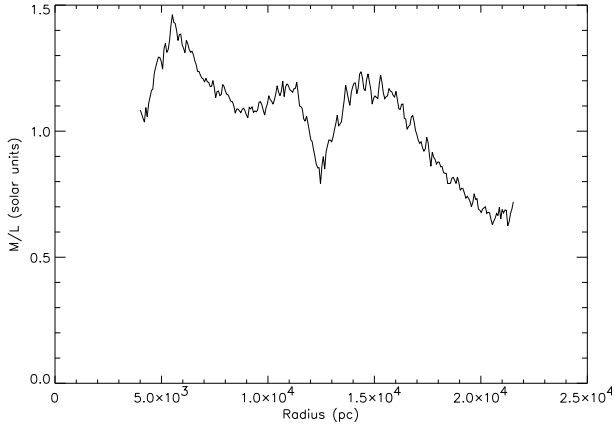,width=85mm}
 \caption{Disc I-band mass-to-light ratio as found by dividing the major-axis luminosity profile into the surface mass. The central 4 kpc has been removed to eliminate the effect of the bulge.}\label{mldisc}
\end{center}
\end{figure}

We calculate the mass-to-light ratio for the disc, initially, by simply dividing the best fit mass profile by the luminosity profile (see Figure \ref{mldisc}). 
The location of a probable spiral arm is visible as a dip at a radius of $\sim$13 kpc. The plot is not flat, as one would expect if the disc light followed a perfect exponential with the scale length used. Instead, there is a noticeable gradient to the ratio. This is possibly an indication of further structure beyond the simple models used. An additional minor difference between our models and those of Schmidt is our use of a thick disc. Finally, the presence of the bar affects the underlying mass distribution. Interestingly, if the disc scale length is increased to $r_d \sim$12kpc, the gradient virtually disappears. This does, however, neglect the bar influence. As a crude estimate of the value of the ratio, however, one can average the result from the curve to find,
\begin{equation}
(M/L)_{I,d} = 1.1\pm0.2h_{70},
\end{equation}
 where $I$ indicates the wavelength band and $d$ denotes the disc. 

This value is consistent with those from the literature. Syer, Mao \& Mo (1998) used disc stability arguments to place an upper bound on the I-band mass-to-light ratio, (M/L)$_I$ $\leq$ 1.9h$_{100}$ $\equiv$ 1.3h$_{70}$. An independent study using population synthesis models (Boissier \& Prantzos, 2000) gives (M/L)$_I$ = 1.0-1.3h$_{100}$ $\equiv$ 0.7-0.9h$_{70}$. Finally, Sommer-Larsen \& Dolgov (2001) use the I-band Tully Fisher relation of Giovanelli et al. (1997) and their warm dark matter simulations to limit the mass-to-light ratio in this band to (M/L)$_I$ = 0.6-0.7h$_{60{\pm}10}$. These independent results fit well with the value we find.

\begin{figure}
\begin{center}
 \epsfig{file=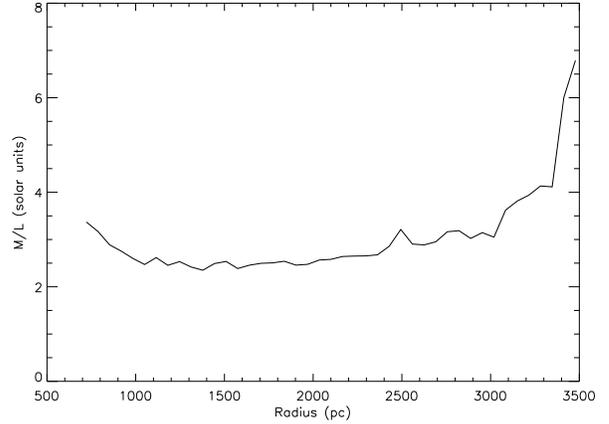,width=85mm}
 \caption{I-band mass-to-light ratio profile for the bulge component. The central 700 pc have been omitted to reduce the effect of contaminating quasar light. The sharp rise at $r\sim$3.3 kpc is the region between the disc and bulge where the mass distributions fail to fit the light profile adequately.}\label{mlbulge}
\end{center}
\end{figure}

By fitting an exponential curve to the light profile, outside of the bulge region of influence, the light contribution of the disc can be removed and the bulge mass-to-light ratio calculated. This technique assumes that the disc light follows the exponential decay to the centre of the galaxy and is not disrupted by the other mass components in that region.  We ignore the light emitted within the image region, as this is contaminated by quasar light. The final profile for the best-fitting bulge model (bg = 9.53 $\times$ 10$^5$ M$_\odot$pc$^{-2}$) is displayed in Figure \ref{mlbulge}.
This profile is reasonably flat over a large range of radii, but increases sharply for $r >$ 3 kpc. This is the beginning of the region where both the disc and bulge are influential. Here, the disc exponential profile is a poor fit to the light profile and the disc subtraction removes too much light. This indicates the inadequacy of the two profiles to accurately account for the mass. It is possible the inclusion of the bar would add mass to the inner regions of the bulge area (its effectiveness is negligible beyond $r\sim$3 kpc) and increase the mass-to-light ratio. This could account for the slope at $r >$ 3 kpc. Unfortunately, as the bulge and bar occupy the same radial regions, they cannot be deconvolved as the disc could. Taking the data from radii, $r$ = 0.7-3.3 kpc, the average mass-to-light ratio is,
\begin{equation}
(M/L)_{I,b} = 2.9\pm0.5h_{70},
\end{equation}
 where $b$ denotes the bulge component. Fukugita, Hogan \& Peebles (1998) find a value of (M/L)$_{B,b}$ = 6.5$\pm^{+1.8}_{-2.0}$ corresponding to (M/L)$_{I,b}$ = 5.0$\pm^{+1.4}_{-1.5}$ in the I-band using synthesis models and data from other groups. This determination just fits with our results. It is encouraging to note the higher mass-to-light ratio in the older population bulge stars than for the younger disc stars.

\subsection{Maximality}\label{max}

Maximal discs have been used in many studies to investigate the dynamics and structure of galaxies, as a means of deconvolving the disc from the dark matter halo. Physically, maximal discs have been found to be inconsistent with  observations of gas motions in spiral galaxies combined with theoretical models (for example, Kranz, Slyz \& Rix 2001).  Observations of spiral galaxies have challenged the suggestion that discs are maximal owing to the need for a significant dark matter halo (e.g. Bottema 1997). Weiner, Sellwood \& Williams (2001) used observed velocity dispersions and gas dynamical simulations to reproduce the observed bar and spiral arms of a large spiral galaxy. They found the disc was maximal (80-100\%) to high confidence. Conversely, Courteau \& Rix (1999) find that their sample HSB galaxies are sub-maximal to high confidence using residuals from the Tully-Fisher relation and adiabatic infall of luminous material into dark matter potentials. Bottema (1993) used stellar velocity dispersions to infer the contribution from stars to the rotation curves in twelve discs. He finds $v_{stars}/v_{total} \sim$ 63$\pm$10 per cent, below maximal according to the definition of Sackett (1997).

The small variation in disc contribution to the 2237+0305 rotation curve, as demonstrated in Figure \ref{rot_curve}, provides a good determination of the degree of maximality in this galaxy. The overall rotation curve, although not constrained observationally in the region where this calculation is made (the disc maximum at $r$ $\sim$ 2.2$r_d$), is also reasonably tight given the HI constraints and the profiles we have used.

The contribution of the disc to the rotation has already been determined by its mean central surface mass density (dc = 506$\pm$30 M$_\odot$pc$^{-2}$). This corresponds to a maximum rotation of $v_{disc}(2.2r_d)$ = 163$\pm$5 kms$^{-1}$. The maximum rotation is calculated to be $v_{total}(2.2r_d)$ = 288$\pm$5 kms$^{-1}$. The percentage contribution of the disc to the rotation, the degree, is therefore,
\begin{equation}
\frac{v_{disc}(2.2r_d)}{v_{total}(2.2r_d)} = 57\pm3 \%
\end{equation}

This value fits well with that found by Bottema (1993), and is well defined for the potential solutions presented in this work. The disc is clearly sub-maximal.

\subsection{Flux Ratios}

The flux observed from each image in a gravitationally lensed system, is a direct measure of the magnification in that region of the lens plane. A comparison between the flux ratios of the images observed in 2237+0305 and those predicted by the solutions can further act as a check on the results.

Observed fluxes in individual images are a combination of magnification due to microlensing, macrolensing and intrinsic variability coupled with time delays. Agol, Jones \& Blaes (2000), however, have measured IR fluxes (8.9 \& 11.7$\mu$m) for the four components and from these the ratio of fluxes can be calculated. In this region of the spectrum, microlensing events are not observed and it is therefore postulated that these observations sample an extended region of the source. In addition, the infrared fluxes are not sensitive to the dust reddening effects of optical light travelling through the galaxy. Thus the IR fluxes should measure the macro-magnification. These fluxes, relative to the B image, are displayed in Table \ref{flux}.

The flux ratios were calculated from our results by taking the ratio of the magnification for each of the images relative to the B image. The magnification is calculated by taking the ratio of areas of triangles around the images mapped from the image to the source plane. They were calculated from the best-fit solution of Table \ref{param_dev} and are displayed in Table \ref{flux}.
\begin{table}
\centering
\begin{tabular}{c|c|c}	\hline
{\em Image} & {\em Flux Ratio (Agol et al.)} & {\em Flux Ratio (this work)}\\	\hline\hline
A & 0.9$\pm$0.1 & 0.7$\pm$0.1\\
B & 1.0 & 1.0\\
C & 0.5$\pm$0.1 & 0.6$\pm$0.1\\
D & 0.9$\pm$0.1 & 1.0$\pm$0.1\\
\end{tabular}\caption{Infrared flux ratios (relative to the B image) of the A, C and D lensed images, calculated from the fluxes of Agol et al. (2000) and our corresponding results.}\label{flux}
\end{table}
Our results are consistent with the observations.

\subsection{Implications for the Dark Matter Halo}

The halo model used is a generic profile that is analytically simple but has little physical motivation. It does, however provide a mass profile with varying slope, and this is a useful attribute if one wishes to study the gradient of the mass distribution.

The best-fit solution for the mass distribution of this galaxy is not an adequate fit. The rotation curve is rising at the outer regions studied instead of falling, and the images are not close enough to their measured points. A halo with a smaller core would alleviate the former problem but require a change of the other mass distributions to address the latter. The mass-to-light profile of the disc suggests an increase in the disc scale length. Such a move would change the shear introduced by the disc. In any case, the core radii found as the best solutions in this study have large core radii, $r_c \sim$ 13-16kpc. These values are not consistent with the cuspy central regions of the CDM profiles. In order to reduce the $\chi^2$ and find an adequate solution, we firstly require further rotation data. The more points we have with good accuracy, the more likely we are to be able to eliminate profiles and find the best solution. Until more data is obtained, it is not worth trying to twist the halo profile to fit the parameters better. If the error bars on the HI rotation points were doubled, the fit would become quite consistent with the observations. The Barnes et al. (1999) observations were undertaken with the VLA C array, a compact configuration with low angular resolution. More informative data is attainable with higher resolution observations.

\begin{figure}
\begin{center}
\epsfig{file=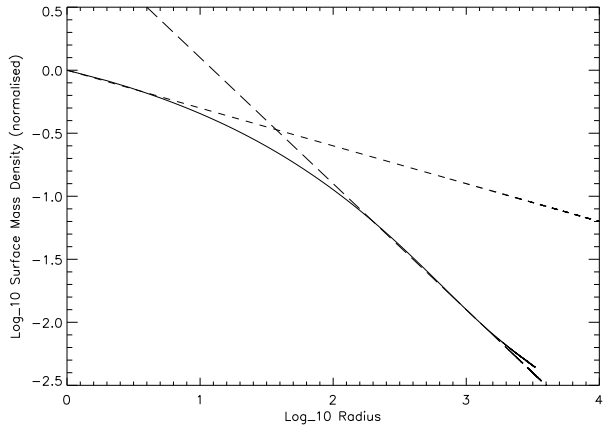,width=85mm}
\caption{Log-log plot of the surface mass density of the inner regions of the best-fit solution (solid line). The short-dashed line indicates the slope of the profile in the central 10pc ($\Sigma \propto r^{-0.3}$). The long-dashed displays the transition to a more isothermal profile ($\Sigma \propto r^{-1}$) outside of the core.}\label{slope}
\end{center}
\end{figure}

The surface mass distribution of the four components combined can be represented on a log-log plot to study the slope as a function of radius. This information is displayed in Figure \ref{slope}.
The dashed lines are normalised fits to $\Sigma \propto r^{-0.3}$ (short-dashed) and $\Sigma \propto r^{-1}$ (long-dashed). The central slope overall is reasonably steep given the large influence of the bulge. The dark matter halo has zero slope in the inner regions given its large core radius. The transition to isothermal is expected given the dominance of the halo in the outer regions. 

\section{Conclusions}

We have undertaken a study of the structure of galaxy 2237+0305 using dynamical and gravitational lensing constraints. The combination of these techniques allows the problems related to the disc/halo degeneracy and maximality of the disc to be overcome.

We find the galaxy is adequately modelled with four mass components, a central bar, stellar disc, bulge and dark matter halo. The bulge contributes $\sim$85 per cent of the convergence of the lensing, within the image radius, and is thus the principal component to which the lensing is sensitive. The disc, bulge and bar all contribute shear, while the assumption of sphericity for the dark matter halo seems adequate for this analysis with the observational information available.

Within the visible radius of the galaxy ($r <$ 30h$_{70}^{-1}$ kpc), the dark matter halo contributes $\sim$60 per cent of the total mass. The potential existence of a core region in the softened isothermal sphere halo is suggested to high confidence by the results. The best reconstruction of the source position occurs for a halo with a non-cuspy central region (within the image radius), and a core radius of 13.4$\pm$0.4kpc, which is equivalent to 1.4$r_d$ for this model.

The disc is found to be sub-maximal to 5$\sigma$ with an average contribution of 57$\pm$3 per cent to the rotation at the maximum of its circular speed. This is a low result, but clearly demonstrates the dominance of the dark matter halo in the galactic mass. The disc and bulge I-band mass-to-light ratios were calculated using the mean and standard deviation of the potential solutions. They were found to be (M/L)$_{I,d}$ = 1.1$\pm$0.2h$_{70}$ and (M/L)$_{I,b}$ = 2.9$\pm$0.5h$_{70}$, respectively. The flux ratios between the quasar images calculated from their magnifications was found to be consistent with infrared data.

With the addition of rotation curve information, the dark matter profile can potentially be found uniquely, and comparisons with the theories of dark matter particle candidates undertaken. With a full rotation curve, the dark matter profile can be found uniquely. The emission lines H$\alpha$, [OIII]$\lambda$5007 and [NII]$\lambda$6583 are all present in the galaxy spectrum (unpublished spectrum from ANU 2.3m, A. Oshlack, private communication) and would be suitable for this study.

\section*{Acknowledgments}
We thank Robert Schmidt for kindly providing his light profiles and for the use of his thesis. CMT acknowledges the funding provided by an Australian Postgraduate Award that has made this work possible, and we thank the anonymous referee for useful comments on the manuscript.

\bsp

\label{lastpage}

\end{document}